\documentclass[article,amsmath,amssymb,a4paper,10pt]{revtex4}
\usepackage{graphicx}
\usepackage{dcolumn}
\usepackage{bm}
\usepackage{psfig}
\usepackage{rotating}
\usepackage{enumerate}
\usepackage{subfigure}
\usepackage{url}
\usepackage[squaren]{SIunits}
\usepackage{textpos}
\usepackage{epstopdf}
\usepackage[hidelinks]{hyperref}
\usepackage{multirow}

\newcommand{\eeZH}{$e^+e^-\rightarrow {ZH}$}
\newcommand{\eevvH}{$e^+e^-\rightarrow\nu\bar\nu {H}$~}
\newcommand{\Htobb}{$H\to b\bar{b}$}
\newcommand{\HtoWW}{$H\to WW^*$}
\newcommand{\HtoZZ}{$H\to ZZ^*$}

\newcommand{\GH}{$\Gamma_H$}
\newcommand{\GZ}{$\Gamma_Z$}
\newcommand{\GW}{$\Gamma_W$}
\newcommand{\BZ}{$\mathrm{BR}_Z$}
\newcommand{\BW}{$\mathrm{BR}_W$}

\begin{document}
\normalsize
\parskip=5pt plus 1pt minus 1pt

\title{Model Independent Determination of $HWW$ coupling and Higgs total width at ILC}

\author{
Claude D$\ddot{\mathrm u}$rig$^a$, Keisuke Fujii$^{b}$, Jenny List$^{a}$, Junping Tian$^{b}$
\\
\vspace{0.2cm}
{\it
$^{a}$ Deutsches Elektronen-Synchrotron (DESY), Hamburg, Germany \\
$^{b}$ High Energy Accelerator Research Organization (KEK), Tsukuba, Japan
}
}


\begin{abstract}
This article is based on the talk presented at the International Workshop on Future Linear Colliders (LCWS13) which held during November 11-15, 2013 at Tokyo, Japan. We present several analyses related to the Higgs total width study at ILC based on the full detector simulation of ILD, which are \eevvH followed by \Htobb~ and \HtoWW. The studies show that at $\unit{250}{GeV}$ we can determine the Higgs total width with a relative precision of 11\% and the $HWW$ coupling with 4.8\%, whereas at $\unit{500}{GeV}$ the expected precision can be significantly improved to 5\% and 1.2\% respectively, assuming the baseline integrated luminosities of ILC, which are $\unit{250}{fb^{-1}}$ @ $\unit{250}{GeV}$, $\unit{500}{fb^{-1}}$ @ $\unit{500}{GeV}$, and a beam polarization of $P(e^{-},e^{+})=(-80\%,+30\%)$. A new approach of removing pile-up particles based on multivariate method developed during those analyses is also presented.
\end{abstract}



\begingroup
\let\newpage\relax

\maketitle
\endgroup


\tableofcontents

\section{Introduction}
Following the discovery of a Standard Model (SM) like Higgs boson with a mass around $\unit{125}{GeV}$, the determination of its total decay width is one of the fundamental physical tasks of investigating its profile. For an SM Higgs of this mass, the expected total width is around $\unit{4}{MeV}$, which is far beyond the detector resolution at both LHC and ILC and therefore it cannot be measured  directly by reconstructing its line shape. So some indirect approaches are proposed in the references \cite{LHCHiggsWidth1,LHCHiggsWidth2,LHCHiggsWidth3,ILCTDRVol2}. Consequently, at LHC, due to the fact that it's impossible to measure the Higgs decay inclusively, the Higgs total width cannot be determined model independently. At ILC, the advantage of recoil mass techniques make the inclusive measurement possible. It measures the absolute cross section of \eeZH, which is proportional to the square of the $HZZ$ coupling ($g^2_Z$). With $g^2_Z$ known, the partial width of \HtoZZ ~(\GZ) ~can be given explicitly. Combined with another measurement of the branching ratio of \HtoZZ ~(\BZ), the Higgs total width (\GH) can be determined by $$\Gamma_H=\frac{\Gamma_Z}{\mathrm{BR}_Z}.$$
In this approach \GZ ~can be measured accurately, however for the SM Higgs the precision of \BZ ~is statistically limited by its small branching ratio \BZ$\sim$2.7\%.  Another approach by utilizing the \HtoWW ~mode, which has a much larger branching ratio \BW$\sim 22\%$, gives $$\Gamma_H=\frac{\Gamma_W}{\mathrm{BR}_W}.$$
The determination of \GW~or $g^2_W$ is not as trivial as in the case of \GZ. To explain the method, let's look at the following five independent observables:
\begin{eqnarray}
	Y_1&=&\sigma_{ZH}=F_1\cdot g^2_{Z}
	\nonumber \\
	Y_2&=&\sigma_{ZH}\times\mathrm{Br}(H\to b\bar{b})=F_2\cdot\frac{g^2_{Z}g^2_{b}}{\Gamma_H}
	\nonumber \\	
	Y_3&=&\sigma_{\nu\bar{\nu}H}\times\mathrm{Br}(H\to b\bar{b})=F_3\cdot\frac{g^2_{W}g^2_{b}}{\Gamma_H}	
	\nonumber \\		
	Y_4&=&\sigma_{\nu\bar{\nu}H}\times\mathrm{Br}(H\to WW^*)=F_4\cdot\frac{g^4_{W}}{\Gamma_H}		
	\nonumber \\
	Y_5&=&\sigma_{ZH}\times\mathrm{Br}(H\to WW^*)=F_5\cdot\frac{g^2_{Z}g^2_{W}}{\Gamma_H},
	\nonumber	
\end{eqnarray}
where $g_{Z},~g_{W}$ and $g_{b}$ are the couplings of Higgs to $ZZ$, $WW$ and $b\bar{b}$ respectively; $F_1,~F_2,~F_3,~F_4$ and $F_5$ are factors which we can calculate unambiguously \cite{HiggsDecayTheory1,HiggsDecayTheory2,HiggsDecayTheory3} (though there are some theory errors either from higher order corrections or from errors of parameters such as $m_H$ or $m_b$, those errors are believed to be well controlled below the sub-percent level). With these five observables the couplings and the total width can be obtained as following:
\begin{enumerate}[i.)]
	\item from the measurement $Y_1$ we can get the coupling $g_{Z}=\sqrt{\frac{Y_1}{F_1}}$.
	\item from the ratio ${Y_2}/{Y_3}$ we can get the coupling ratio $g_{Z}/g_{W}=\sqrt{\frac{Y_2}{Y_3}\frac{F_3}{F_2}}$.
	\item with $g_{Z}$ and $g_{Z}/g_{W}$, we can get $g_{W}=\sqrt{\frac{Y_1Y_3}{Y_2}\frac{F_2}{F_1F_3}}$.
	\item {\bf option A}: once we know $g_{W}$, from the measurement $Y_4$ we can get the Higgs total width $\Gamma_H=\frac{Y_1^2Y_3^2}{Y_2^2Y_4}\frac{F_2^2F_4}{F_1^2F_3^2}$;  {\bf option B}: once we know $g_Z$ and $g_W$, from the measurement $Y_5$ we can get the Higgs total width $\Gamma_H=\frac{Y_1^2Y_3}{Y_2Y_5}\frac{F_2F_5}{F_1^2F_3}$.
\end{enumerate}
These two options are constrained by both the Higgs-strahlung \eeZH~and the WW-fusion \eevvH production processes. 
At $\unit{250}{GeV}$, the former one reaches its maximum cross section, but latter has a small cross section, thus making option B the more suitable method. The main limiting factor arises from the precision of the measurement $Y_3$. Going up to $\unit{500}{GeV}$, the WW-fusion production cross section is around one order larger compared to $\unit{250}{GeV}$, which makes option A the better option. It is worth emphasizing that eventually the precisions of $2\frac{\Delta Y_1}{Y_1}$ and $\frac{\Delta Y_4}{Y_4}$ limit the precision of the total width, since $Y_2$  usually is far better measured than $Y_1$, and $Y_3$ twice better than $Y_4$.

In this article, we focus on the analyses of measuring the observables $Y_3$ and $Y_4$ through the WW-fusion channel. The analyses for $Y_1,~Y_2$ and $Y_3$ through the Higgs-strahlung channel have been investigated in \cite{RecoilMass,ZHtobb,ZHtoWW}. At both $\unit{250}{GeV}$ and $\unit{500}{GeV}$, $Y_3=\sigma_{\nu\bar{\nu}H}\times\mathrm{Br}(H\to b\bar{b})$ is studied since it is mandatory in option A and option B; $Y_4=\sigma_{\nu\bar{\nu}H}\times\mathrm{Br}(H\to WW^*)$ is studied only at $\unit{500}{GeV}$ since it is useless at $\unit{250}{GeV}$, and both the hadronic and semi-leptonic decay of $WW^*$ are investigated.

Nevertheless, the determination of deviation of $HWW$ coupling to its SM value itself has important impact on the existence of additional Higgs boson \cite{HWWTheory1,HWWTheory2}. In the $W_LW_L\rightarrow W_LW_L$ scattering process, the unitarity is insured by the $HWW$ coupling which is constrained to be $2m_W^2/v$ in SM. If deviation smaller to its SM value is found, it strongly indicates that there exists another neutral heavier Higgs boson which also contributes to the scattering process. Depending on the size of deviation, if it's around 1\%, it suggests the new heavier Higgs mass would be lighter than 10 TeV, and if it's around 10\%, the new heavier Higgs mass would be lighter than 2 TeV. However, if it turns out that deviation larger than SM value is found, interesting enough is the indication that there exists double charged new Higgs boson with similar scale constraints as the formal deviation case. So the precision measurement of $HWW$ coupling would provide strong indication of next energy scale we need explore. 

\section{Simulation Framework}
All the signal and background samples for 500 GeV analyses are generated using the common DBD softwares \cite{DBDSoftwares}, and LoI softwares \cite{LoISoftwares} for 250 GeV analysis, based on the full detector simulation of ILD by GEANT4. WHIZARD \cite{WHIZARD} is used as event generator, detector simulation is done by Mokka \cite{Mokka}, and reconstruction is done by Marlin \cite{Marlin}. Particle flow is carried out by PandoraPFA \cite{PandoraPFA}, package LCFIPlus \cite{LCFIPlus} and LCFIVertex\cite{LCFIVertex} are used to provide flavor tagging information.

\section{Analysis of $\sigma_{\nu\bar{\nu}H}\times\mathrm{Br}(H\to b\bar{b})$ @ 250 GeV}
The feasibility of the measurement of the Higgs production cross section through WW-fusion is investigated for $\sqrt{s}=$ 250 $\mathrm{GeV}$ and a beam polarization of $P(e^{+}e^{-}) =(0.3,-0.8)$, assuming $\unit{250}{fb^{-1}}$ of data.
We can extract information on the coupling $g_{W}$ of the Higgs boson to W-bosons which then provides us the possibility to determine the total decay width of the Higgs boson. The SM Higgs boson with a mass below $\unit{140}{GeV}$ is expected to decay predominantly into two b-quarks
\[
{e^+}{e^-}\longrightarrow \nu_{{e}}\bar{\nu}_{{e}}{H}\longrightarrow \nu_{{e}}\bar{\nu}_{{e}}{b\bar{b}}\,.
\]
In the ${H}\nu_{e}\bar{\nu}_{e}$ final state, Higgs-strahlung and WW-fusion cannot be taken separately as non-interfering. The WW-fusion cross-section increases logarithmically to large $\sqrt{s}$, whereas Higgs-strahlung scales as $s^{-1}$. Hence and due to the enhanced cross section at the threshold $\sqrt{s}= m_{\rm H} + m_{\rm Z}$ Higgs-strahlung diagrams give the dominant contribution to the combined process at low energies and thus representing one of the most challenging backgrounds in the analysis. Next to Higgs-strahlung, backgrounds considered in this search mode are two-fermion events, having a production cross-section which is more than 1000 times larger than the signal cross-section, semi-leptonically and hadronically decaying Z/W-pairs, and single Z/W-boson production processes. Background events from two-photon processes are found to be negligible. The cross-section of the two-photon interaction is very large, but their particular features allow to suppress them at an early stage of the analysis. The backgrounds are divided into the following types: ${b}{\bar{b}}\nu_{l}\bar{\nu}_{l}$, ${q}{\bar{q}}\nu_{l}\bar{\nu}_{l}$ (${q}\neq{b}$), ${q}{\bar{q}}\nu_{l}{l}$, ${q}{\bar{q}}{l}^{-}{l}^{+}$, ${q}{\bar{q}}{q}{\bar{q}}$ and ${q}{\bar{q}}$. 
The large background contribution around $\sqrt{s}=\unit{250}{GeV}$ make the analysis very challenging.
The signal search mode consists of missing four-momentum and two energetic, very forward b-jets. In the beginning of the analysis all reconstructed particles are clustered into two jets representing the Higgs decay products. The event selection is performed in three stages. The first step involves pre-cuts, using the number of charged tracks $N_{\rm ctrk}$ and the removal of isolated leptons in the events. Since the neutrino mode is selected, there are no signal events containing isolated leptons. In the signal leptons appear in the jets at most. Removing events with isolated leptons leads to a reduction of semi-leptonic backgrounds ${q\bar{q}l^{+}l^{-}}$ and ${q\bar{q}l}\nu_{l}$. Background events originating from hadronic decays of W- and Z-pairs, as well as $q\bar{q}$-pairs, can contain more particles compared to the WW-fusion signal. The introduced limits on the number of charged tracks in an event $10 \leq N_{\rm ctrk} \leq 40$ help to reduce ${q\bar{q}l^{+}l^{-}}$, ${q\bar{q}q\bar{q}}$ and ${q\bar{q}}$ background and exclude fully leptonic events for sure. The second stage of the event selection contains cuts on kinematic variables and in the third stage more event specific cuts are performed, mainly using jet characteristics and variables resulting from jet clustering and flavor tagging. In the following, the selection cuts are discussed briefly:
\begin{itemize}
\item Cut1: visible mass has to be consistent with $m_{\rm H} - \unit{20}{GeV} \leq m_{\rm{vis}} \leq m_{\rm H} + \unit{10}{GeV}$. It has a great effect on the ${q\bar{q}}$-background since ISR photons, which are preferably emitted in beam-pipe direction and might escape detection faking missing energy, can bring the invariant visible mass of the two-fermion system back to $m_{\rm{vis}}\approx m_{\rm Z} $. Even though the invariant mass of the ${q\bar{q}}$-background peaks at $m_{\rm Z}$, the tail of the visible mass distribution is still large, leaving this the dominant background. Additionally, W- or Z-pair background get rejected well.
\item Cut2: visible energy is required to be $\unit{105}{GeV} \leq E_{\rm{vis}} \leq\unit{160}{GeV}$. This selection cut is not very effective, mainly reducing ${q\bar{q}}$-background.
\item Cut3: absolute value of visible and invisible transverse momentum $p_{\rm{T}}$ are equal due to momentum conservation. In backgrounds without neutrinos (${q\bar{q}l^{+}l^{-}}$, ${q\bar{q}q\bar{q}}$, ${q\bar{q}}$) missing $p_{\rm{T}}$ can be caused by particles that stay undetected. Hence, those backgrounds mainly consist of events with low $p_{\rm{T}}$. 
A requirement on the total transverse momentum $\unit{20}{GeV} \leq \sum p_{\rm{T}} \leq\unit{80}{GeV}$ reduces backgrounds without neutrinos in the final state.
 \item Cut4 \& Cut5: the Durham jet clustering algorithm offers two variables $Y_{23}$ and $Y_{12}$. These parameters are useful to discriminate between events with different numbers of jets. Events have to satisfy $Y_{23}\leq 0.02$, which is the threshold value to reconstruct two jets as three jets. 
To further discriminate between signal and background, a cut on the second parameter is applied $0.2 \leq Y_{12} \leq0.8$, which corresponds to the minimum $Y$-parameter at which the number of jets changes from two to one for the two-jet hypothesis.
\item Cut6: flavor tagging is performed by using the LCFIVertex flavor tagging package. It is based on a neutral net approach to distinguish b-, c- and light jets. The b-jet likelihood should fulfill $\rm{btag} \geq 0.85$.  
Due to the near absence of b-quarks in backgrounds originating from W- and Z-bosons, these processes can be reduced.
\item Cut7: total jets momentum in beam direction should satisfy $|\sum p_{\rm{z}}| \leq\unit{60}{GeV}$. Since the W- and Z-boson in the corresponding backgrounds are relatively boosted, $p_{\rm{z}}$ is larger as compared to WW-fusion and Higgs-strahlung events. The cut is very helpful to reduce a large part of the two-fermion background contribution. 
\item Cut8: Z- and W-bosons are produced at small angles from the ${e^{+}e^{-}}$-beams and therefore the angular distribution of these processes have peaks in the forward and backward regions. A cut on $| \cos(\theta_{\rm jet}) | \leq 0.95$ is applied.
\end{itemize} 
During the event selection, more cuts have been tested to further reduce background, without the desired effect. The Higgs-strahlung and WW-fusion distribution of the different cut parameters are most of the time of similar shape thus making the choice of cuts less effective for Higgs-strahlung events. The effect of each cut on signal and background events are listed in table~\ref{tab:vvHbb250}. 
\begin{table}[htbp]
\small
\caption{The reduction table for the signal and backgrounds in the analysis of $\nu\bar{\nu}H\to\nu\bar{\nu}b\bar{b}$ at $\unit{250}{GeV}$. The cut names are explained in the text. $\nu\bar{\nu}H$ is divided into WW-fusion and Higgs-strahlung.}
\label{tab:vvHbb250}
\centering
\begin{tabular*}{0.8\textwidth}{@{\extracolsep{\fill}}l|r|r|r|r|r|r|r|r|r|r}
   \hline Process & expected & pre-selection & Cut1 & Cut2 & Cut3 & Cut4 & Cut5 & Cut6 & Cut7 & Cut8\\
   \hline \hline
   $\nu\bar{\nu}H (\mathrm{fusion})$ & 3426 & 2663 & 2070 & 2023 & 1577 & 1053 & 965 & 547 & 519 & 507 \\ \hline
   $\nu\bar{\nu}H (ZH)$ & $1.4\times 10^{4}$ & 10918 & 8356 & 8356 & 7448 & 4860 & 4594 & 2574 & 2546 & 2546 \\ \hline   
   $\nu_{l}\bar{\nu}_{l}{b\bar{b}}$ & $3.05\times 10^{4}$ & 23012 & 1040 & 1040 & 878 & 421 & 390 & 224 & 193 & 187 \\ \hline      
   $\nu_{l}\bar{\nu}_{l}{q\bar{q}}$ & $1.19\times 10^{5}$ & 88998 & 5548 & 5545 & 4714 & 2408 & 2271 & 15 & 9 & 9 \\ \hline         
   ${q\bar{q}l^{+}l^{-}}$ & $2.99\times 10^{5}$ & 153540 & 6196 & 5922 & 1760 & 588 & 508 & 65 & 38 & 36 \\ \hline            
   ${q\bar{q}l}\nu$ & $1.73\times10^6$ & $1.15\times10^6$ & 181973 & 177193 & 134047 & 22654 & 20533 & 111 & 73 & 65\\ \hline               
   ${q\bar{q}q\bar{q}}$ & $3.91\times10^6$ & $1.15\times10^6$& 782 & 728 & 3 & 1 & 0 & 0 & 0 & 0 \\ \hline                  
   ${q\bar{q}}$ & $26.02\times10^6$ & $17.27\times10^6$ & 852321 & 794892 & 1507 & 1199 & 683 & 289 & 152 & 152 \\ \hline                                     
   \hline
   $\mathrm{BG}$ & $32.104\times10^6$ & $19.846\times10^6$ & $1.047\times10^6$  & 985320 & 142909 & 27271 & 24385 & 1404 & 465 & 449 \\ \hline                                                       
\end{tabular*}
\end{table}

After the selection, the dominant background to WW-fusion is represented by Higgs-strahlung. The remaining background contribution is in the same order as the signal. In order to determine the WW-fusion cross section, we modify the relation $\sigma_{\nu\bar{\nu}H}({H}\rightarrow{b\bar{b}}) = \sigma_{\nu\bar{\nu}H}\times BR({H}\rightarrow{b\bar{b}})$ to
\begin{equation}
\sigma_{\nu\bar{\nu}H}\times BR({H}\rightarrow{b\bar{b}}) = \frac{N'_{\rm{WW}}}{\epsilon\cdot\mathcal{L}}\,,
\label{eq:wwfusioncross}
\end{equation}
where $\epsilon$ is the selection efficiency and $\mathcal{L}$ the integrated luminosity. 
It follows, that by extracting the number of WW-fusion events $N'_{\rm WW}$ which have passed the event selection, $\sigma_{\nu\bar{\nu}H}\times BR({H}\rightarrow{b\bar{b}})$ can be determined.
The WW-fusion events with $\nu\bar{\nu}{b\bar{b}}$ final state can be separated from the corresponding one in Higgs-strahlung by exploiting their different characteristics in the $\nu\bar{\nu}$ invariant mass, which are measurable through the missing mass distribution. Therefore, a $\chi^{2}$-fit is applied on the shape of the missing mass distribution consisting of the remaining WW-fusion, Higgs-strahlung and background events and by using Toy Monte Carlo data as reference.
In a $\chi^{2}$-fit, the function
\begin{equation}\nonumber
\chi^{2} = \sum_{i}^{N_{\rm bins}}(N_{i}^{\rm pred} - N_{i}^{\rm data})^{2} / \sigma^{2}(N_{i}^{\rm pred}),
\label{eq:chi}
\end{equation}
has to be minimized, where $N_{i}^{\rm data}$ and $N_{i}^{\rm pred}$ is the number of data and predicted events in bin $i$. In order to fit on the missing mass distribution consisting of background, Higgs-strahlung and WW-fusion, we need to set up $N_{i}^{\rm pred}$ as a function of the three distributions:
\begin{equation}\nonumber
N_{i}^{\rm pred} = f_{\rm{WW}}N'_{{\rm WW},i} + f_{\rm{ZH}}N'_{{\rm ZH},i} + f_{{\rm bgrd}}N'^{\rm tot}_{{\rm bgrd},i},
\end{equation}
where $N'_{{\rm WW},i}$, $N'_{{\rm ZH},i}$ and $N'^{\rm tot}_{{\rm bgrd},i}$ represent the number of events in bin $i$ after the selection, respectively. The parameters $f_{\rm{WW}}, f_{\rm{ZH}}$ and $f_{\rm{bgrd}}$ are adjusted so as to minimize the $\chi^{2}$-function. 
\begin{figure}[ht]
  \centering
    \includegraphics[width=0.5\columnwidth]{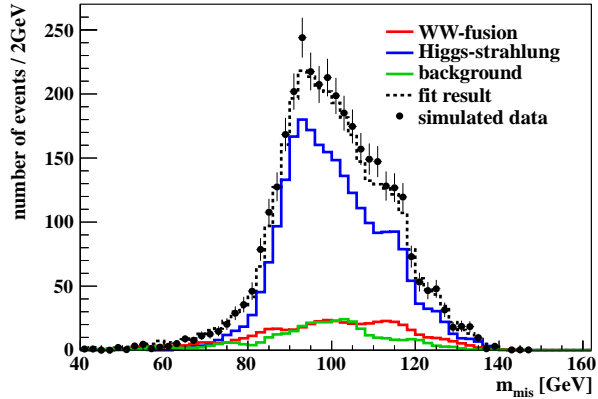}
  \caption{Missing mass distribution of WW-fusion, Higgs-strahlung and background for $m_{\rm H}=\unit{125}{GeV}$ after cuts, including the fit result. The shape of the Higgs-strahlung distribution is expected to peak at $m_{\rm Z}$, whereas WW-fusion is expected to peak at slightly larger missing masses for $\unit{250}{GeV}$. Latter is quasi-flat due to the small number of WW-fusion events.}
  \label{fig:vvHbb250}
\end{figure}
\\
The result of the fit yields the cross-section $\sigma_{\nu\bar{\nu}H}({H}\rightarrow{b\bar{b}})$, since only ${H}\rightarrow{b\bar{b}}$ decays are selected. This is a simplified assumption assuming only true b-jet particles have passed the event selection. The fit result depicted in figure~\ref{fig:vvHbb250} states as results:
\begin{table}[h]
\centering
 \begin{tabular}{l|r|r|r}
 \hline
$\text{Process}$ &  $N'_{\rm WW}\pm\Delta N'_{\rm WW}$  & $N'_{\rm ZH}\pm\Delta N'_{\rm ZH}$ & $N'^{\rm tot}_{\rm bgrd}\pm\Delta N'^{\rm tot}_{\rm bgrd}$  \\
\hline
$\text{Fit result}$ & $512\pm 54$ & $2\,497\pm 85$ & $454\pm 46$\\\hline
\end{tabular}
\end{table}
\\
 The relative precision of $\sigma_{\nu\bar{\nu}H}\times BR({H}\rightarrow{b\bar{b}})$ can be determined by using gaussian error propagation of Equation~\ref{eq:wwfusioncross}.
The uncertainty in the efficiency is considered negligible. Systematic effects of the luminosity are not considered in the analysis. 
Taking into account the uncertainties from the fit and from the branching ratio measurement, the precision of $\sigma_{\nu\bar{\nu}H}\times BR({H}\rightarrow{b\bar{b}})$ is expected to be $10.5\,\%$.\\

\section{Analysis of $\sigma_{\nu\bar{\nu}H}\times\mathrm{Br}(H\to b\bar{b})$ @ 500 GeV}

The analysis of this mode at $\unit{500}{GeV}$ is quite similar to the one at $\unit{250}{GeV}$, except that the cross section of \eevvH through WW-fusion is almost one order larger at $\unit{500}{GeV}$, $\sim$ $\unit{150}{{fb}}$. High statistics of signal events which are due to the large cross section and the large branching ratio of \Htobb ~, offer the opportunity of a precision measurement. The signal final state consists of two missing neutrinos and two b-jets. For the pre-selection, it is natural to reconstruct two jets from all reconstructed particles, and to reject events with isolated charged leptons. This efficiently suppresses the backgrounds such as those including leptonic decays of W or Z. Each event, either signal or background, is overlaid with beam induced $\gamma\gamma\to\mathrm{hadrons}$ events~\cite{Overlay}. 
The cross section of this overlay increases significantly as the center-of-mass energy rises. At $\unit{500}{GeV}$, an average of 1.7 $\gamma\gamma\to\mathrm{hadrons}$ events per bunch crossing is estimated. So before using the inclusive jet clustering algorithm, some methods are used to remove the overlaid particles from those we are interested in.

\begin{figure}[ht]
  \centering
  \begin{tabular}[c]{ccc}
    \includegraphics[width=0.3\columnwidth]{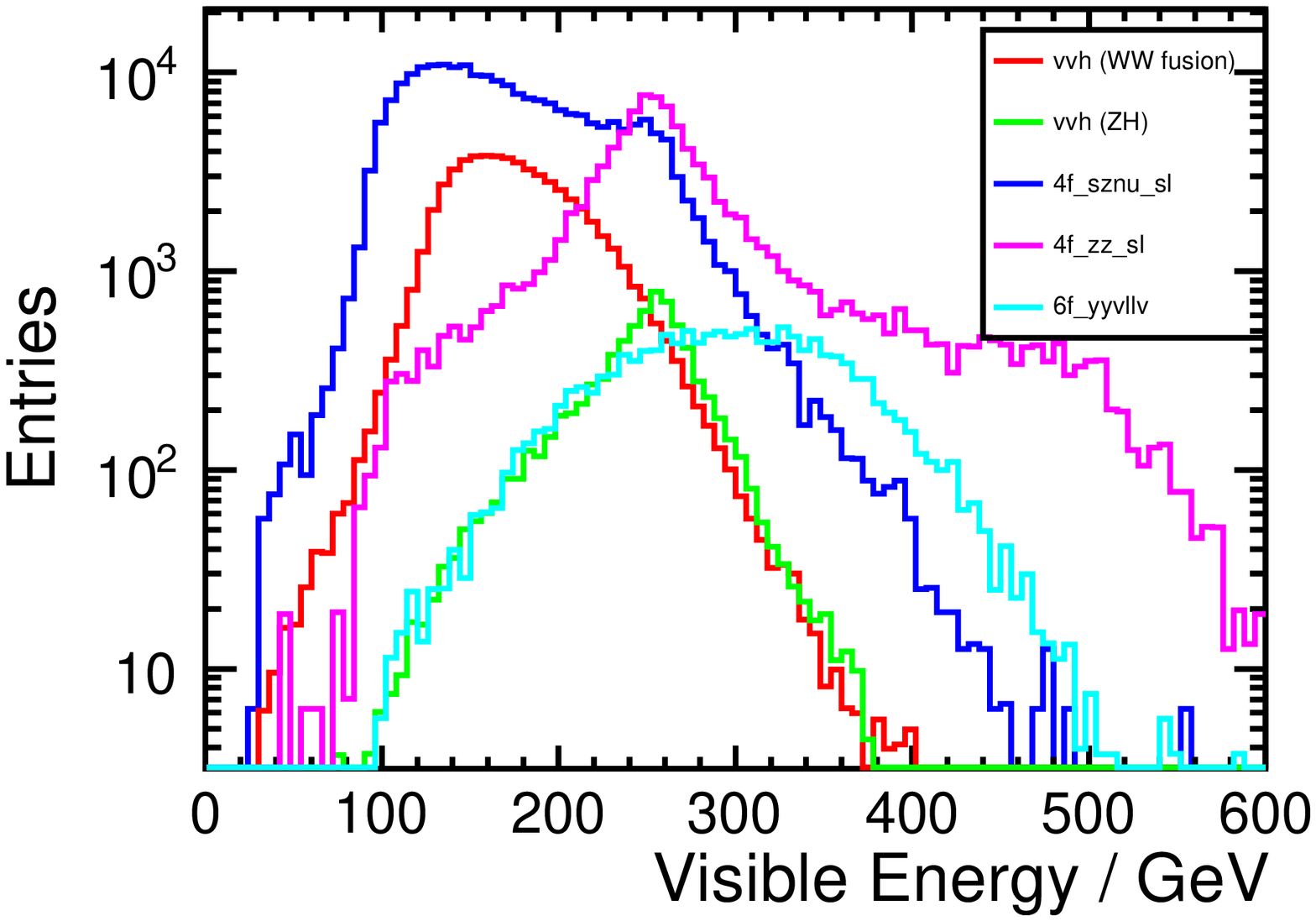}  &
    \includegraphics[width=0.3\columnwidth]{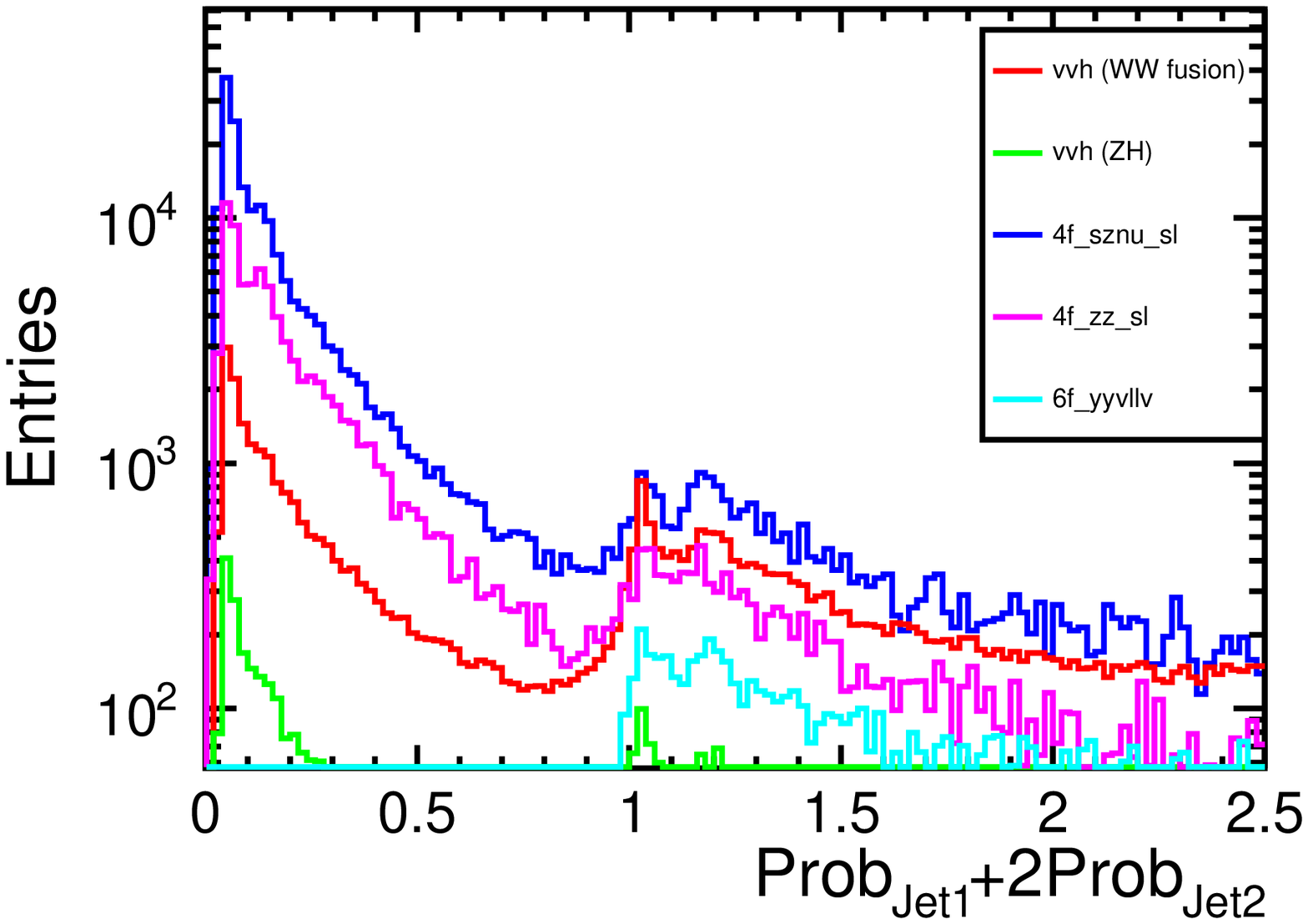} &
    \includegraphics[width=0.3\columnwidth]{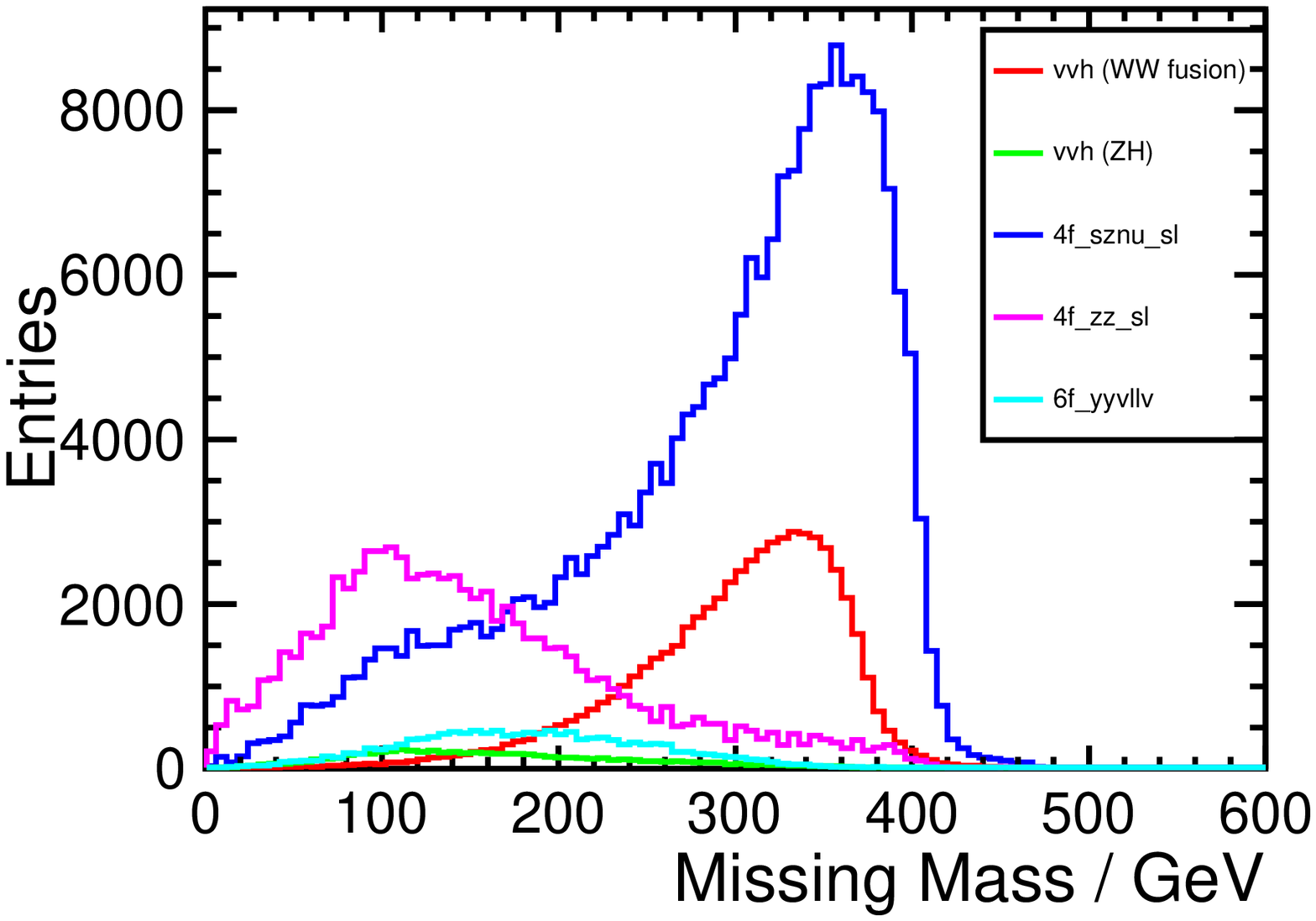} \\
 \end{tabular}
  \caption{Distributions of visible energy (left), b-likeness (middle) and missing mass (right) for signal $\nu\bar{\nu}H$ and backgrounds, where $\mathrm{4f\_sznu\_sl}$ is for 4-fermions from $\nu\bar{\nu}Z\to\nu\bar{\nu}qq$, $\mathrm{4f\_zz\_sl}$ is for 4-fermions from $ZZ\to\nu\bar{\nu}qq$ and $\mathrm{6f\_yyvllv}$ is mainly 6-fermions from the leptonic decay of $t\bar{t}$.}
  \label{fig:vvHbb500Vars}
\end{figure}

\begin{figure}[ht]
  \centering
    \includegraphics[width=0.5\columnwidth]{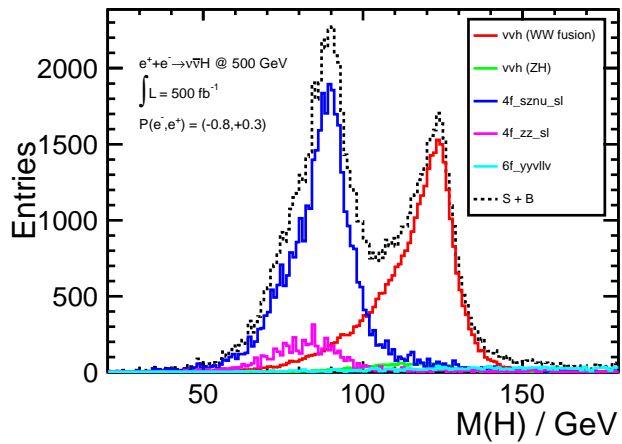}
  \caption{Distribution of the reconstructed Higgs invariant mass using \Htobb.}
  \label{fig:vvHbb500Higgs}
\end{figure}

The dominant background processes considered in this analysis are 4-fermion processes from $e^++e^-\to \nu\bar{\nu}Z,~ZZ,~e\nu W, ~W^+W^-$, and 6-fermion processes mainly from $e^++e^-\to t\bar{t}$. For the final selection, we require large missing energies and missing Pt to significantly suppress the fully hadronic backgrounds. The reconstructed jets need to be tagged as b-jets which significantly suppresses the light-quark jet backgrounds. In order to separate the signal contribution from \eeZH , a missing mass larger than the Z-boson mass is required. Distributions of those related variables after pre-selection for both signal and backgrounds are plotted in figure~\ref{fig:vvHbb500Vars}. After the final selection, figure~\ref{fig:vvHbb500Higgs} gives the distribution of the reconstructed Higgs invariant mass, where both the Higgs peak from the signal and the Z peak from $\nu\bar{\nu}Z$ can be seen clearly. Eventually, a cut on the Higgs invariant mass is applied to suppress the $\nu\bar{\nu}Z$ background.

\subsection{Removal of Overlay}
\begin{figure}[ht]
  \centering
  \begin{tabular}[c]{cc}
    \includegraphics[width=0.45\columnwidth]{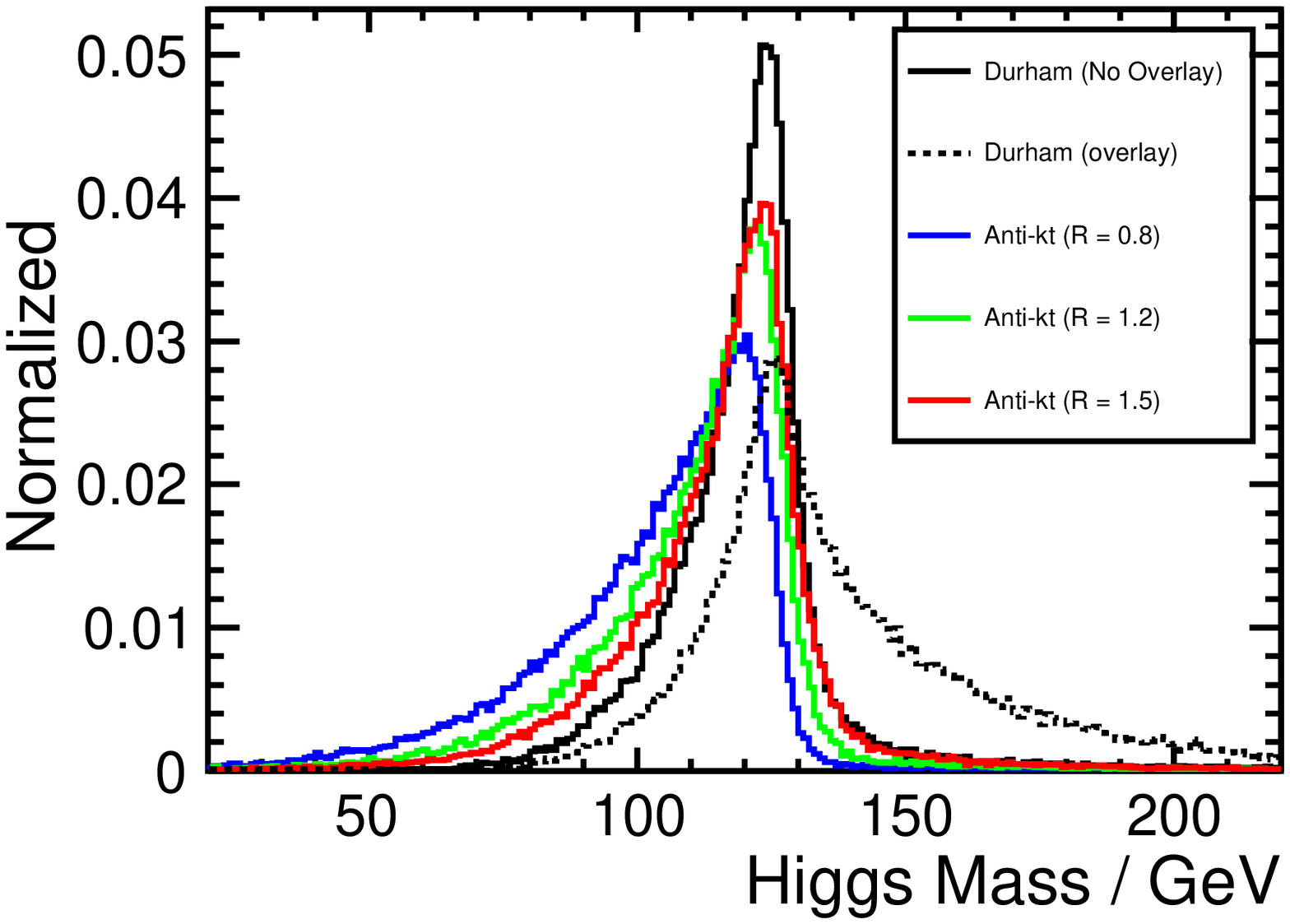}  &
    \includegraphics[width=0.45\columnwidth]{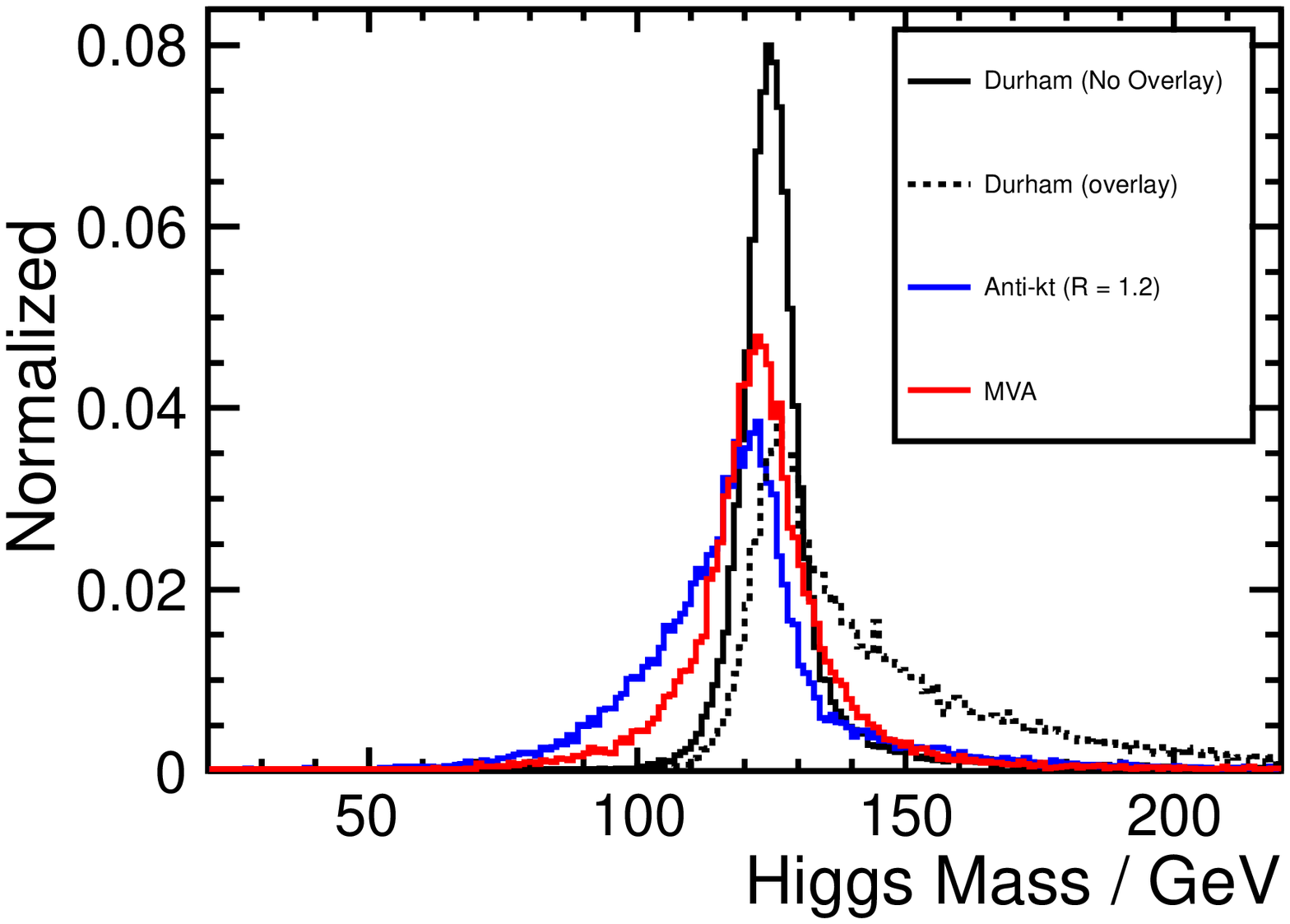}  \\
 \end{tabular}
  \caption{Comparison of the reconstructed Higgs invariant mass by different options to remove the overlay in case of \Htobb ~(left) and \HtoWW ~(right).}
  \label{fig:OverlayMass}
\end{figure}

Two methods are developed to remove particles originating from the overlaid beam background. One of them is based on the $k_{\mathrm{T}}$ or anti-$k_{\mathrm{T}}$ jet clustering algorithm. The overlaid particles usually have very low Pt and a large polar angle in the forward and backward regions which is very close to beam direction. 
This usually makes distances defined in the $k_{\mathrm{T}}$ or anti-$k_{\mathrm{T}}$ algorithm between overlaid particles and particles from the target process very large. Hence the beam background particles can be effectively un-selected by the jet clustering. The R value is optimized to get the best Higgs mass resolution after the overlay removal, which is shown in figure~\ref{fig:OverlayMass} (left). This method usually works well for target processes with high Pt hard jets, for instance \Htobb ~in this analysis where R is optimized to be 1.5.

\begin{figure}[ht]
  \centering
  \begin{tabular}[c]{ccc}
    \includegraphics[width=0.3\columnwidth]{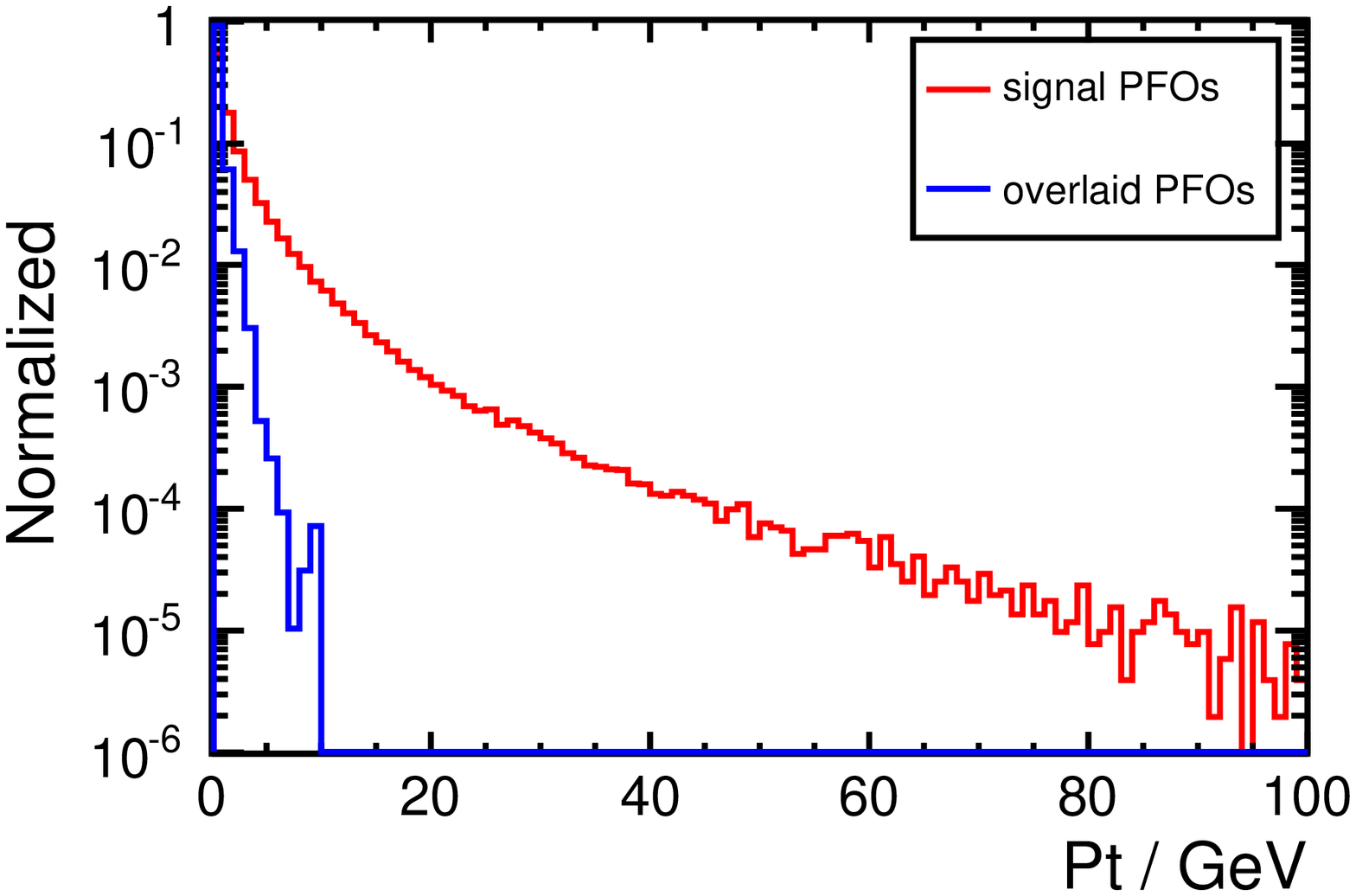}  &
    \includegraphics[width=0.3\columnwidth]{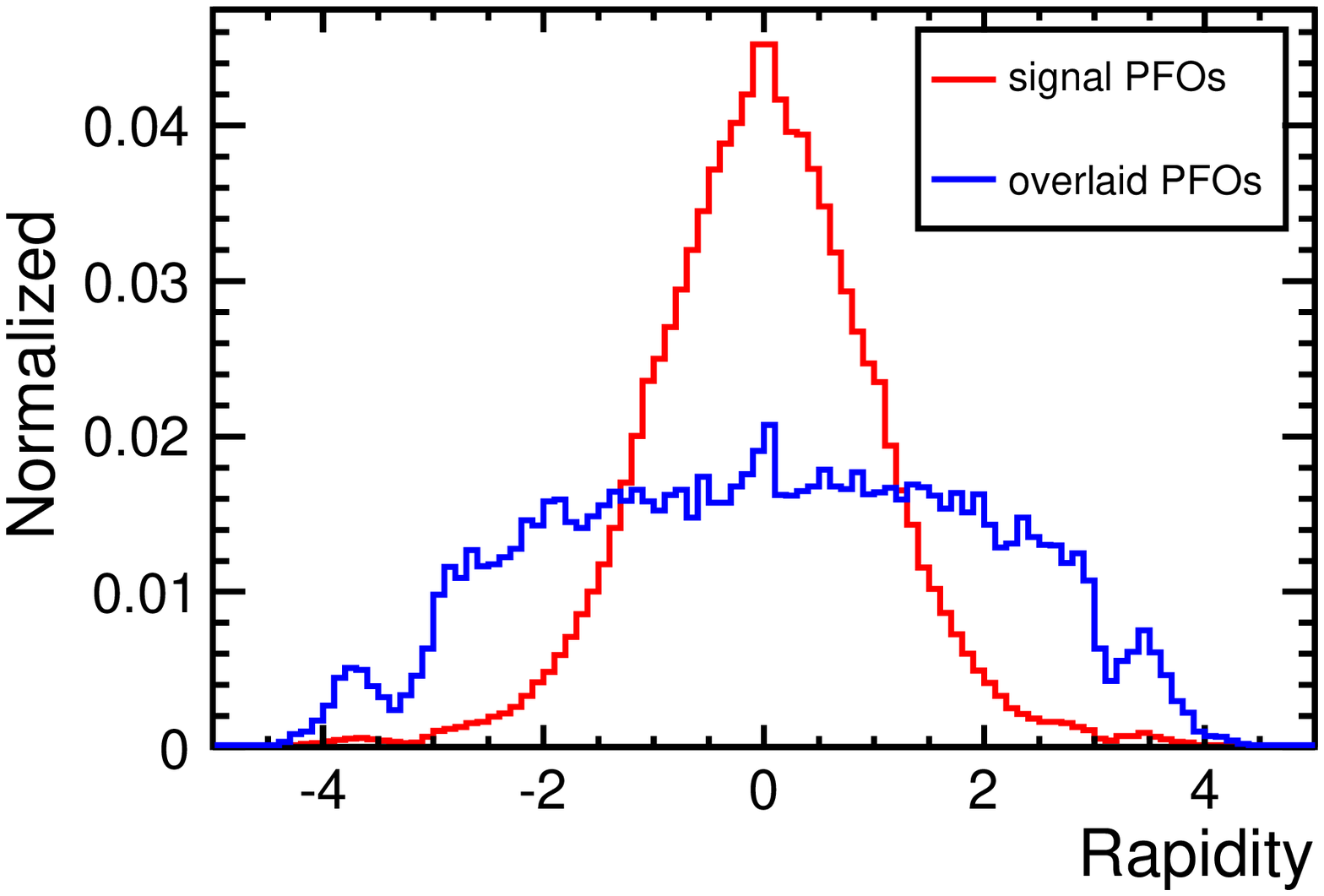} &
    \includegraphics[width=0.3\columnwidth]{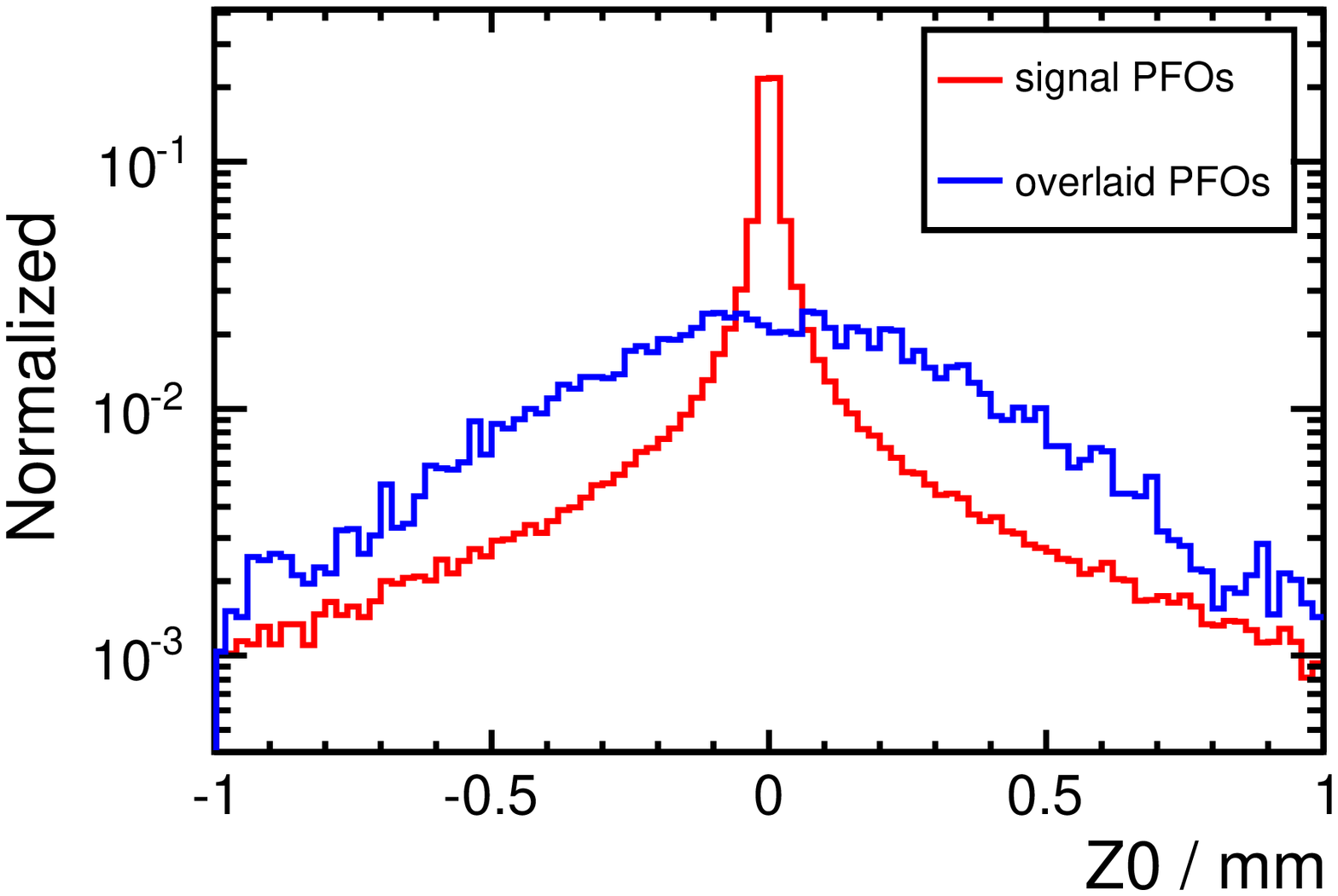} \\    
 \end{tabular}
  \caption{Distributions of Pt (left), rapidity (middle) and $z_0$ of IP (right, only for charged) for particles from overlay process or target process.}
  \label{fig:OverlayVars}
\end{figure}

\begin{figure}[ht]
  \centering
  \begin{tabular}[c]{cc}
    \includegraphics[width=0.4\columnwidth]{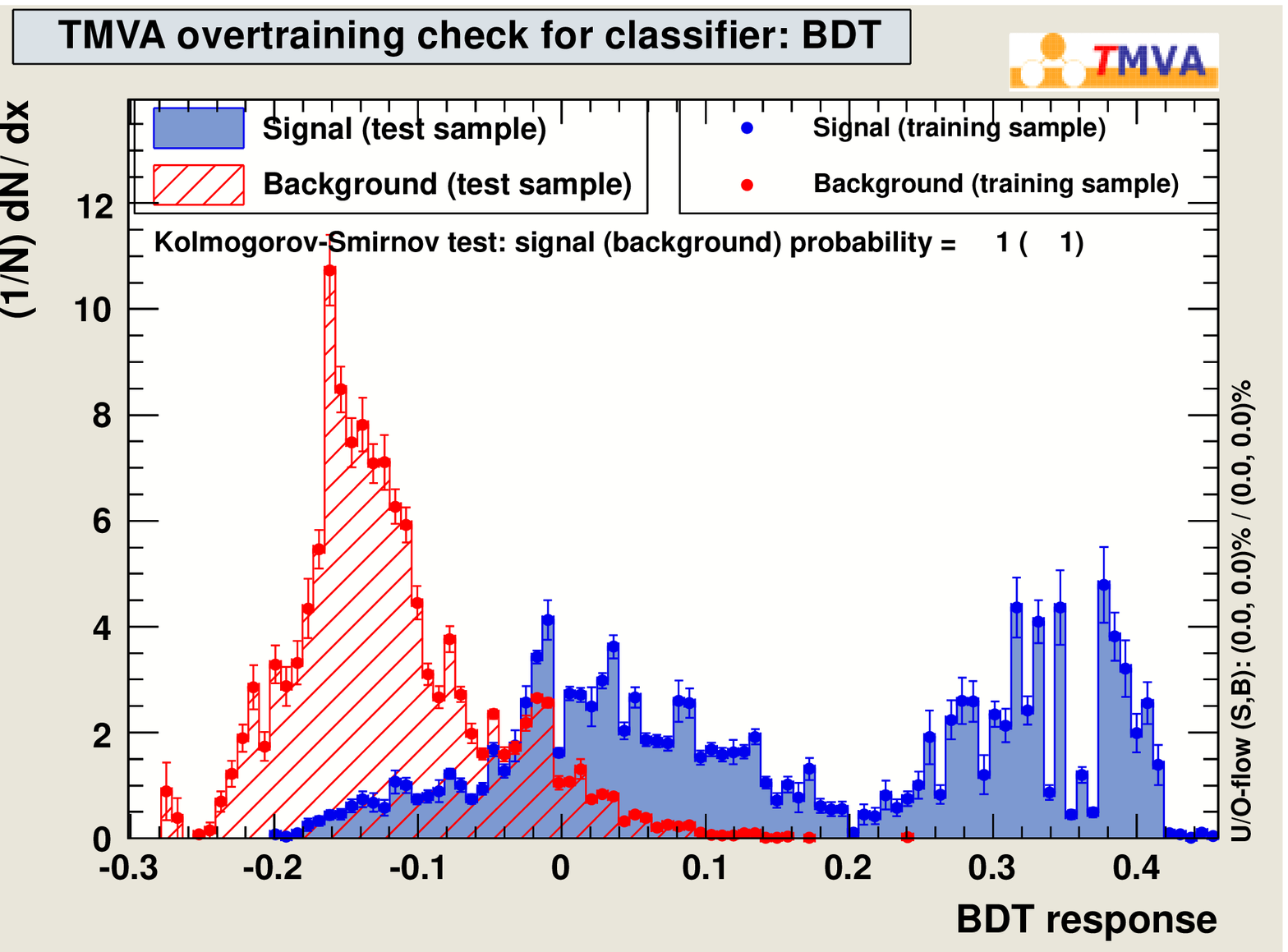}  &
    \includegraphics[width=0.4\columnwidth]{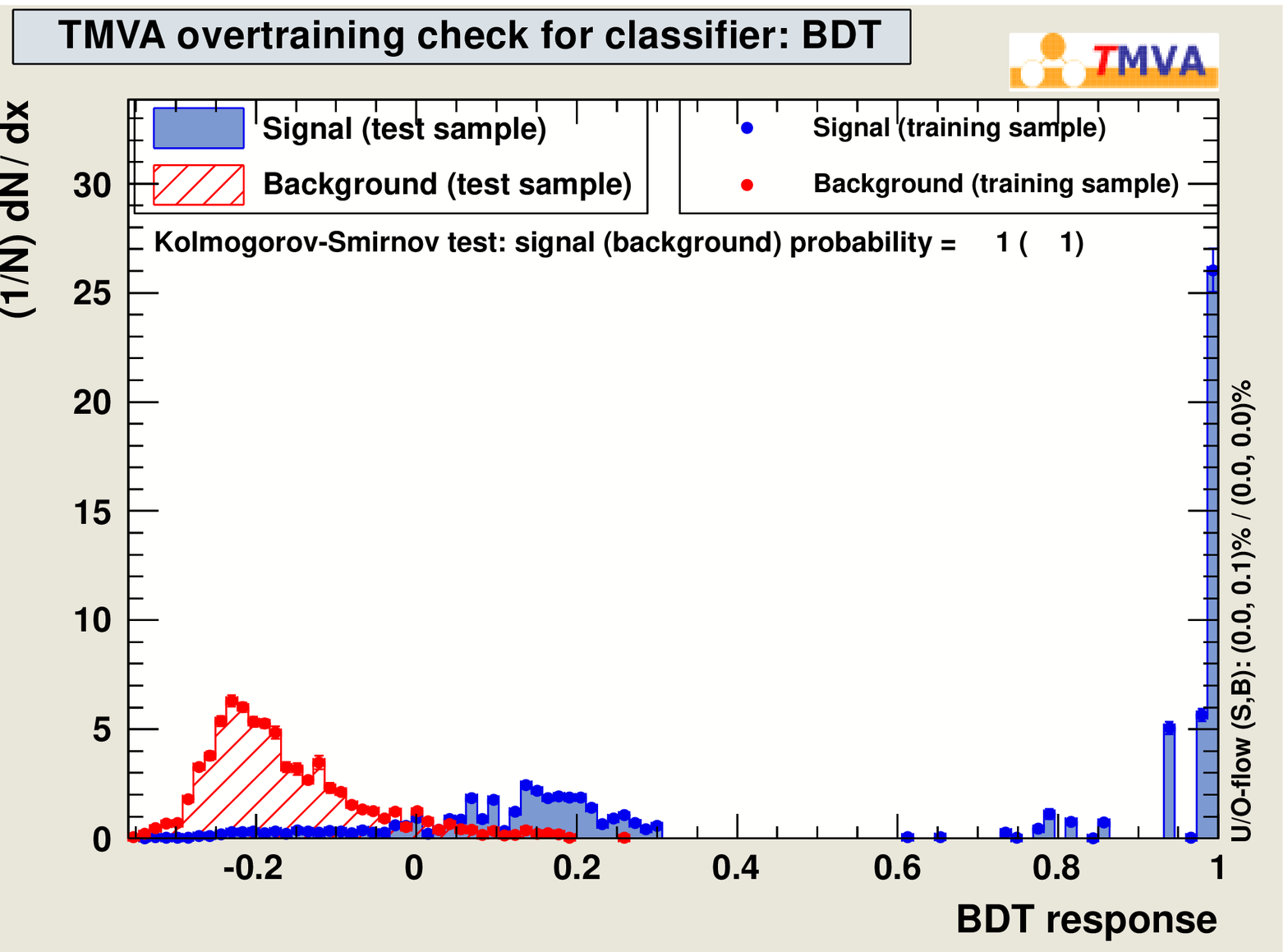}  \\
 \end{tabular}
  \caption{BDT Output for two categories: neutral particles (left) and charged particles (right).}
  \label{fig:OverlayBDT}
\end{figure}

Instead of using the method based on jet clustering, another particle based approach can be used, in which the overlaid particles are tagged one by one from the information on Pt and the polar angle. In addition, the IP information can be utilized to tag the overlaid particles, since the IP of the overlay process can have some sizable shift to the IP of the target process. With these information, a multivariate method, BDT here, is implemented to give a likeness of being overlay for each particle. The variables used in BDT are shown in figure~\ref{fig:OverlayVars}. The BDT is trained for two categories, neural particles and charged particles. In the former one only Pt and rapidity are used and $z_0$ is added to the later one. As stated before, the output of the BDT depicts the likeness of events being beam background overlay, which is shown in figure~\ref{fig:OverlayBDT}. A relatively large likeness of overlay is required to tag the overlaid particles. This particle based method shows better performance than the jet clustering based method in case of relatively soft jets, such as jets from $W^*$ in the analysis of \HtoWW, as shown in figure~\ref{fig:OverlayMass} (right). 

\subsection{Event selection and reduction table}
The following steps are carried out orderly in the pre-selection:
\begin{itemize}
	\item The anti-$k_T$ jet clustering based method is applied to remove the beam induced overlay with $\mathrm{R}=1.5$, which is implemented by the package FastJetClustering.
	\item An event is rejected if any isolated charged lepton is found.
	\item After the overlay removal, the remaining particles are clustered into two jets, each of which is flavor tagged. This step is implemented by the package LCFIPlus.
	\item Each jet is required to have at least 8 particles.
\end{itemize}

The following cuts are applied orderly in the final selection:
\begin{itemize}
	\item Cut1: the visible energy is required to be smaller than $\unit{300}{GeV}$ but larger than $\unit{100}{GeV}$, and the total Pt is required to be larger than $\unit{20}{GeV}$.
	\item Cut2: the charged lepton with largest momentum ($P(Lmax)$) in the remaining particles is required to have relatively larger cone energy ($E_{cone}$), which is $P(Lmax)<2E_{cone}+\unit{20}{GeV}$.
	\item Cut3: the b-likenesses of the two jets ($Prob(Jet1)$, $Prob(Jet2)$) are required to be large, $Prob(Jet1)+2Prob(Jet2)>0.92$.
	\item Cut4: the missing mass is required to be larger than $\unit{172}{GeV}$.
	\item Cut5: the reconstructed Higgs invariant mass is required to be larger than $\unit{100}{GeV}$ but smaller than $\unit{143}{GeV}$.
\end{itemize}

The remaining numbers of signal and background events after each cut are shown in the reduction table~\ref{tab:vvHbb500}. Eventually, assuming an integrated luminosity of $\unit{500}{{fb}^{-1}}$ and a beam polarization of $P(e^{-},e^{+})=(-80\%,+30\%)$, 29199 signal events of which 28598 are from \Htobb ,~and 7176 background events dominated by $\nu\bar{\nu}Z$ and $ZH$ are selected. The statistical significance is $150\sigma$ and the relative precision of $\sigma_{\nu\bar{\nu}H}\times\mathrm{Br}(H\to b\bar{b})$ at $\unit{500}{GeV}$ is expected to be $0.667\%$, which is consistent with what extrapolated from LoI results in DBD of $0.661\%$.

\begin{table}[htbp]
\caption{The reduction table for signal and backgrounds in the analysis of $\nu\bar{\nu}H\to\nu\bar{\nu}b\bar{b}$ at $\unit{500}{GeV}$. The cut names are explained in text. $\nu\bar{\nu}H$ has two types, one of signal WW-fusion process, the other from ZH process. The number of signal events after Cut5 in the parenthesis is for \Htobb.}
\label{tab:vvHbb500}
\centering
\begin{tabular*}{0.8\textwidth}{@{\extracolsep{\fill}}l|r|r|r|r|r|r|r}
   \hline Process & expected & pre-selection & Cut1 & Cut2 & Cut3 & Cut4 & Cut5\\
   \hline \hline
   $\nu\bar{\nu}H (\mathrm{fusion})$ & $7.47\times10^4$ & 59698 & 54529 & 54048 & 35598 & 34278 & 299199 (28598) \\ \hline
   $\nu\bar{\nu}H (ZH)$ & $1.02\times10^4$ & 7839 & 7301 & 7224 & 4863 & 1951 & 1512 \\ \hline   
   $\mathrm{4f\_sznu\_sl}$ & $2.79\times10^5$ & 234259 & 203489 & 202977 & 44943 & 39125 & 3957 \\ \hline      
   $\mathrm{4f\_sw\_sl}$ & $2.43\times10^6$ & 228436 & 135164 & 121791 & 1495 & 911 & 132 \\ \hline         
   $\mathrm{4f\_zz\_sl}$ & $1.83\times10^5$ & 102172 & 60684 & 59865 & 13036 & 5736 & 461 \\ \hline            
   $\mathrm{4f\_ww\_sl}$ & $2.78\times10^6$ & 653997 & 287428 & 250944 & 3851 & 1145 & 176 \\ \hline               
   $\mathrm{4f\_sze\_sl}$ & $9.41\times10^5$ & 65011 & 1311 & 1259 & 91.1 & 40.7 & 5.51 \\ \hline                  
   $\mathrm{6f\_yyveev}$ & $6.05\times10^3$ & 931 & 306 & 104 & 96.6 & 87.4 & 20.4 \\ \hline                     
   $\mathrm{6f\_yyvelv}$ & $2.37\times10^4$ & 5450 & 2425 & 1116 & 997 & 907 & 237 \\ \hline                        
   $\mathrm{6f\_yyvllv}$ & $2.36\times10^4$ & 8009 & 4272 & 2813 & 2556 & 2383 & 674 \\ \hline                           
   \hline
   $\mathrm{BG}$ & $6.68\times10^6$ & $1.31\times10^6$ & 702379 & 648094 & 71929 & 52285 & 7176 \\ \hline                              
   $\mathrm{significance}$ & 16.6 & 35.0 & 43.3 & 44.6 & 106 & 114 & 150 \\ \hline                                 
\end{tabular*}
\end{table}

\section{Analysis of $\sigma_{\nu\bar{\nu}H}\times\mathrm{Br}(H\to WW^*)$ @ 500 GeV}
Depending on the decay mode of each W, two analyses are carried focusing on full hadronic and semi-leptonic decays of $WW^*$.
\subsection{$WW^*\to 4\textnormal{-}\mathrm{jets}$}
In this mode, the final state consists of two missing neutrinos and four jets none of which is a b-jet. In the pre-selection, it is essential to reconstruct the four jets and to pair them according to one on-shell $W$ and one off-shell $W^*$. Then the Higgs mass can be fully reconstructed. The main background processes considered here are similar to those in the \Htobb ~analysis, dominated by $\nu\bar{\nu}Z$, $e\nu W$ and $W^+W^-$. 

The following steps are carried out orderly in the pre-selection:
\begin{itemize}
	\item The particles based method is applied to remove the beam background overlay, which is implemented according to the method in IV-A.
	\item An event is rejected if any isolated charged lepton is found.
	\item After the overlay removal, remaining particles are clustered into four jets, each of which is flavor tagged. This step is implemented by the package LCFIPlus.
	\item Due to the jets originating from off-shell $W^*$, the number of particles in each jet, which are ordered by the energy from largest to smallest, is required to be no smaller than 7, 6, 5, 4, and in total the number of particles need be no smaller than 40.
\end{itemize}

The following cuts are applied orderly in the final selection:
\begin{itemize}
	\item Cut1: the Y values obtained from jet clustering should match the features of four jet events in order to suppress backgrounds with fewer partons such as $\nu\bar{\nu}Z\to\nu\bar{\nu}qq$, $e\nu W\to e\nu qq$, or $W^+W^-\to \nu\bar{\nu}qq$. Nevertheless, one should keep in mind that no perfect jet clustering algorithm here can reject all the two partons events. The Y values are required to satisfy $Y_{4\to3}>0.0026$ and $Y_{3\to2}>0.0076$.
	\item Cut2: the visible energy is required to be smaller than $\unit{230}{GeV}$, total Pt is required to be larger than $\unit{20}{GeV}$ and the missing mass is required to be larger than $\unit{200}{GeV}$, which significantly suppresses the contribution from ZH.
	\item Cut3: the charged lepton with largest momentum ($P(Lmax)$) in the remaining particles is required to have relatively larger cone energy ($E_{cone}$), which is $P(Lmax)<2E_{cone}+\unit{9}{GeV}$.
	\item Cut4: to suppress events with b-jets, such as \Htobb, $t\bar{t}$, etc., the b-likenesses of the four jets (sorted from largest to smallest $btag1$, $btag2$, $btag3$, $btag4$) are required to satisfy $btag1+2btag2<0.7$ and $btag3+2btag4<0.14$.
	\item Cut5: the reconstructed on-shell $W$ mass is required to be larger than $\unit{54}{GeV}$ but smaller than $\unit{94}{GeV}$, the off-shell $W^*$ mass is required to be smaller than $\unit{64}{GeV}$ but larger than $\unit{11}{GeV}$.
	\item Cut6: the reconstructed Higgs invariant mass is required to be larger than $\unit{114}{GeV}$ but smaller than $\unit{142}{GeV}$.
\end{itemize}

\begin{figure}[ht]
  \centering
    \includegraphics[width=0.5\columnwidth]{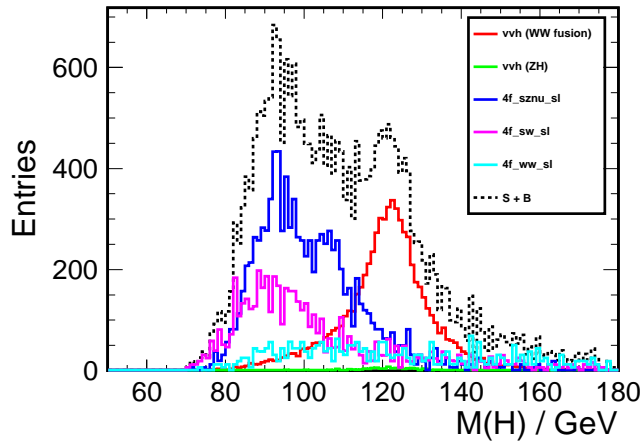}
  \caption{Distribution of the reconstructed Higgs invariant mass using the fully hadronic mode of \HtoWW.}
  \label{fig:vvHWW4j500Higgs}
\end{figure}

The reconstructed Higgs invariant mass after the first five cuts is depicted in figure~\ref{fig:vvHWW4j500Higgs}. The remaining numbers of signal and background events after each cut are listed in the reduction table~\ref{tab:vvHWW4j500}. Eventually, by assuming an integrated luminosity of $\unit{500}{{fb}^{-1}}$ and a beam polarization of $P(e^{-},e^{+})=(-80\%,+30\%)$, 4945 signal events of which 3136 are from \HtoWW ~and 3055 background events dominated by $\nu\bar{\nu}Z$, $e\nu W$ and $W^+W^-$ are selected. The statistical significance is $35\sigma$ and the relative precision of $\sigma_{\nu\bar{\nu}H}\times\mathrm{Br}(H\to WW^*)$ is expected to be $2.8\%$.

\begin{table}[htbp]
\caption{The reduction table for the signal and backgrounds in the analysis of $\nu\bar{\nu}H\to\nu\bar{\nu}WW^*\to\nu\bar{\nu}+4\textnormal{-}\mathrm{jets}$ at $\unit{500}{GeV}$. The cut names are explained in text. $\nu\bar{\nu}H$ has two types, one of signal WW-fusion process, the other from ZH process. The number of signal events after Cut6 in the parenthesis is for \HtoWW.}
\label{tab:vvHWW4j500}
\centering
\begin{tabular*}{0.8\textwidth}{@{\extracolsep{\fill}}l|r|r|r|r|r|r|r|r}
   \hline Process & expected & pre-selection & Cut1 & Cut2 & Cut3 & Cut4 & Cut5 & Cut6\\
   \hline \hline
   $\nu\bar{\nu}H (\mathrm{fusion})$ & $7.47\times10^4$ & 42373 & 14461 & 11684  & 11315 & 7415 & 6746 & 4970(3136) \\ \hline
   $\nu\bar{\nu}H (ZH)$ & $1.02\times10^4$ & 5497 & 911 & 240 & 232 & 144 & 120 & 86.8 \\ \hline   
   $\mathrm{4f\_sznu\_sl}$ & $2.79\times10^5$ & 140092 & 23016 & 18123 & 17841 & 14157 & 9675 & 1308 \\ \hline      
   $\mathrm{4f\_sw\_sl}$ & $2.43\times10^6$ & 220670 & 40715 & 11746 & 11383 & 11013 & 5317 & 778 \\ \hline         
   $\mathrm{4f\_zz\_sl}$ & $1.83\times10^5$ & 57640 & 7041 & 722 & 690 & 546 & 342 & 65.1 \\ \hline            
   $\mathrm{4f\_ww\_sl}$ & $2.78\times10^6$ & 416386 & 46390 & 4816 & 4149 & 3934 & 2965 & 806 \\ \hline               
   $\mathrm{4f\_sze\_sl}$ & $9.41\times10^5$ & 45911 & 19160 & 38.4 & 38.4 & 32.1 & 8.56 & 0 \\ \hline                  
   $\mathrm{6f\_yyveev}$ & $6.05\times10^3$ & 52.5 & 35.7 & 9.24 & 0.02 & 0 & 0 & 0 \\ \hline                     
   $\mathrm{6f\_yyvelv}$ & $2.37\times10^4$ & 703 & 498 & 102 & 45.6 & 9.51 & 5.78 & 3.88 \\ \hline                        
   $\mathrm{6f\_yyvllv}$ & $2.36\times10^4$ & 2025 & 1420 & 358 & 252 & 30.4 & 26.6 & 7.60 \\ \hline                           
   \hline
   $\mathrm{BG}$ & $6.68\times10^6$ & $8.89\times10^5$ & 139185 & 36156 & 34632 & 29866 & 18462 & 3055 \\ \hline                              
   $\mathrm{significance}$ & 3.0 & 6.8 & 13.4 & 19.4 & 19.5 & 21.0 & 24.6 & 35.0 \\ \hline                                 
\end{tabular*}
\end{table}

\subsection{$WW^*\to l\nu+2\textnormal{-}\mathrm{jets}$}
In this mode, the final state consists of three missing neutrinos, one isolated charged lepton and two jets. In the pre-selection, it is essential to find the isolated charged lepton and to reconstruct the two jets. One $W$ can be fully reconstructed from the two jets. However the other $W$ cannot be fully reconstructed from the isolated charged lepton and one missing neutrino due to the other two missing neutrinos originating from $WW$-fusion. Hence, the Higgs mass cannot be fully reconstructed either. The main background processes considered here are similar to those in the $WW^*\to 4\textnormal{-}\mathrm{jets}$~analysis, dominated by $W^+W^-\to l\nu qq$. 

The following steps are carried out orderly in the pre-selection:
\begin{itemize}
	\item Select events with one isolated electron or muon from all particles, otherwise the event is rejected.
	\item The particles based method is applied to remove the beam background overlay, which is implemented according to the method used in IV-A.
	\item The remaining particles are clustered into two jets, each of which is flavor tagged. This step is implemented by the package LCFIPlus.
	\item To suppress events in which the reconstructed jets are actually $\tau$-jets, each jet is required to have at least two charged particles with relatively high Pt ($>\unit{500}{MeV}$).
\end{itemize}

The following strategies are used in the final selection:
\begin{itemize}
	\item depending on the type of the selected isolated charged lepton, all events are separated into two categories, electron-type or muon-type. This is due to background contamination and hence the cut optimization is very different for these two categories.
	\item large missing energy and large missing Pt are required to suppress the fully hadronic backgrounds.
	\item the b-likenesses of the two jets are required to be small to suppress backgrounds with b-jets.
	\item to suppress the dominant background $W^+W^-\to l\nu qq$, the angle between reconstructed $W$ from two jets and isolated charged lepton is required to be relatively small.
	\item  in case of the electron-type category, the polar angle of the electron is required to be not close to beam direction to suppress the $eeZ$ and $e\nu W$ backgrounds; the angle between electron and each of the two jets is required to be relatively large since the selected electron can be a mis-identified electron from the jets.
	\item the partially reconstructed Higgs invariant mass ($m(lqq)$) is still useful to further suppress the backgrounds.
\end{itemize}

\begin{figure}[ht]
  \centering
  \begin{tabular}[c]{cc}
    \includegraphics[width=0.45\columnwidth]{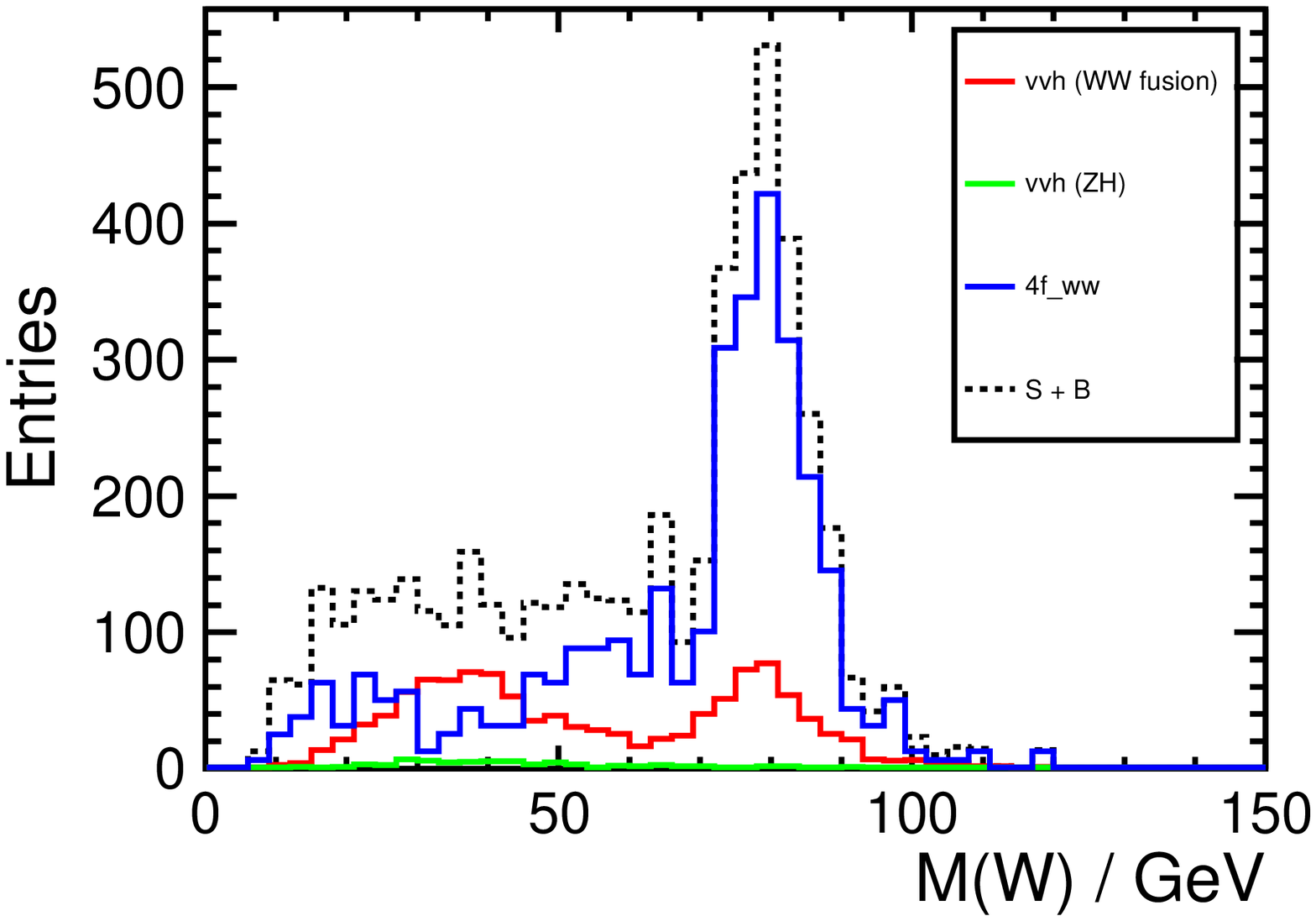} &
    \includegraphics[width=0.45\columnwidth]{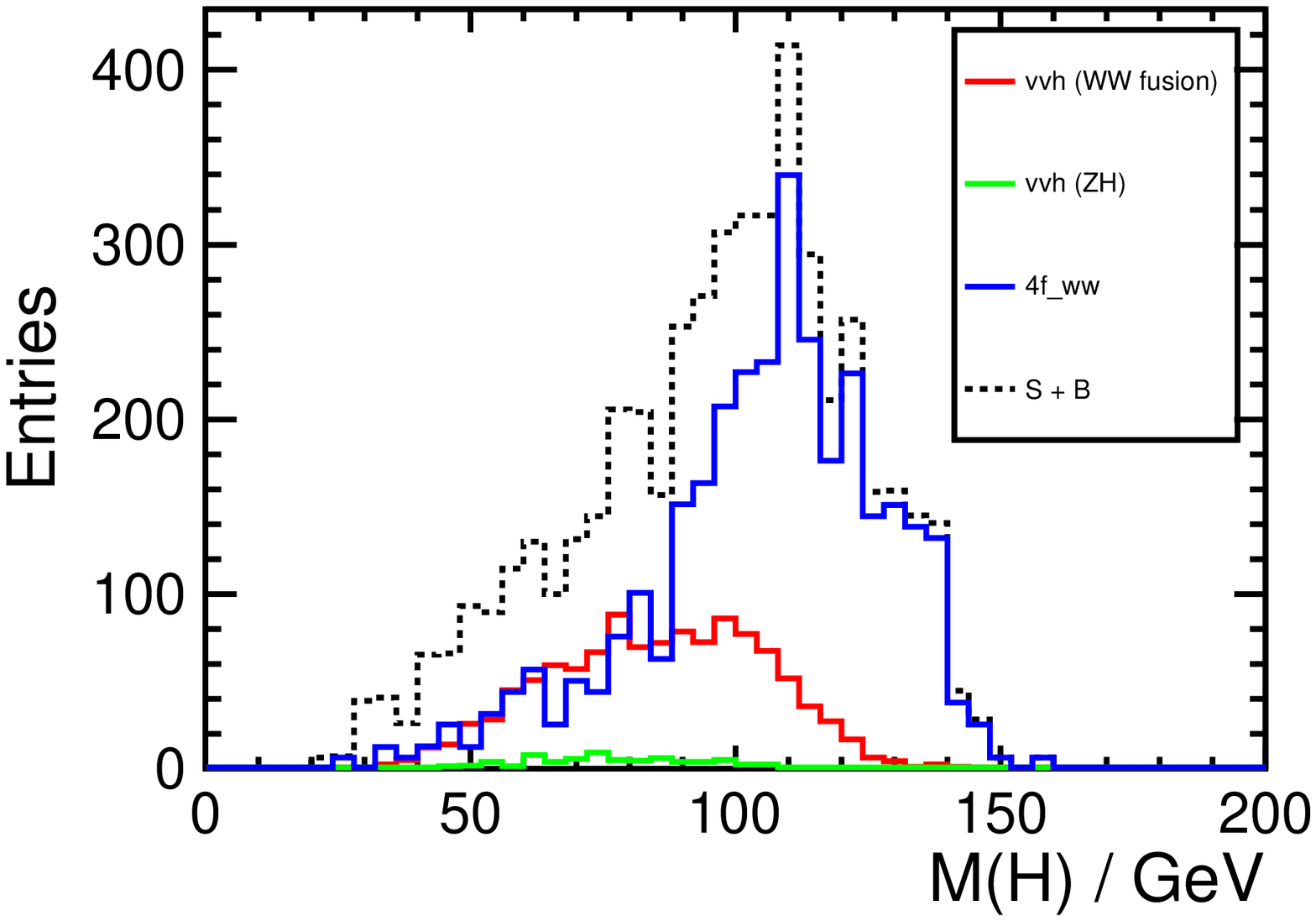}    \\
  \end{tabular} 
  \caption{Distribution of fully reconstructed $W$ mass (left) and partially reconstructed Higgs mass (right) using the semi-leptonic mode \HtoWW.}
  \label{fig:vvHWWlv2j500}
\end{figure}

The fully reconstructed $W$ mass from the two jets, as well as the partially reconstructed Higgs mass from the lepton and two jets are shown in figure~\ref{fig:vvHWWlv2j500}. The remaining numbers of signal and background events after all cuts are shown in the reduction table~\ref{tab:vvHWWlv2j500} for both categories. Eventually, by assuming an integrated luminosity of $\unit{500}{{fb}^{-1}}$ and a beam polarization of $P(e^{-},e^{+})=(-80\%,+30\%)$, the statistical significance for the muon-type category is $17.4\sigma$ and $14.7\sigma$ for the electron-type. The combined result is $22.8\sigma$. The relative precision of $\sigma_{\nu\bar{\nu}H}\times\mathrm{Br}(H\to WW^*)$ is expected to be $4.4\%$ using the semi-leptonic decay mode of $WW^*$.

\begin{table}[htbp]
\caption{The remaining numbers of signal and background events for the two categories in the analysis of $\nu\bar{\nu}H\to\nu\bar{\nu}WW^*\to\nu\bar{\nu}+l\nu+2\textnormal{-}\mathrm{jets}$ at $\unit{500}{GeV}$. The number of signal events in the parenthesis is for \HtoWW.}
\label{tab:vvHWWlv2j500}
\centering
\begin{tabular*}{0.5\textwidth}{@{\extracolsep{\fill}}|l|r|r|r|}
   \hline category & signal & background & significance \\ \hline
   muon-type & 1002 (982) & 2187 & $17.4\sigma$ \\ \hline
   electron-type & 879 (858) & 2528 & $14.7\sigma$ \\ \hline   
\end{tabular*}
\end{table}

\subsection{Combined of \HtoWW ~with full hadronic and semi-leptonic modes}
By combining the two decay modes of $WW^*$, the relative precision of $\sigma_{\nu\bar{\nu}H}\times\mathrm{Br}(H\to WW^*)$ at $\unit{500}{GeV}$ is expected to be $2.4\%$, assuming an integrated luminosity of $\unit{500}{{fb}^{-1}}$ and a beam polarization of $P(e^{-},e^{+})=(-80\%,+30\%)$. 

\section{Summary}
The relative precision of $\sigma_{\nu\bar{\nu}H}\times\mathrm{Br}(H\to b\bar{b})$ at $\unit{250}{GeV}$ is expected to be $10.5\%$ with 250 $\mathrm{fb}^{-1}$ data, and at $\unit{500}{GeV}$ is expected to be $0.667\%$ with 500 $\mathrm{fb}^{-1}$ data, assuming beam polarisations $P(e^{-},e^{+})=(-80\%,+30\%)$ at both energies. The Higgs total width is expected to be measured with precision of 13\%  at 250 GeV according to Option B, and 5.4\% at 500 GeV according to Option A. By adding \HtoZZ ~and other decay modes \cite{GlobalFit}, the expected precision of Higgs total width at 250 GeV only is 11\%, and by combining 250 GeV and 500 GeV data is 5.0\%. $HWW$ coupling can be determined to a precision of 4.8\% at 250 GeV and 1.2\% at 500 GeV with the baseline luminosities of ILC. The results are summarized in Table \ref{tab:HWWGamma}, where the expectations with luminosity upgrade scenarios \cite{LumiUP} of ILC are also shown. The capability of sub-percent level measurement of $HWW$ coupling will be crucial to hint at next energy scale of new physics beyond SM. 

\begin{table}[t]
 \begin{center}
 \begin{tabular}{|l|r|r|r|r|}
  \hline
  \multirow{2}{*}{${\Delta g}/{g}$} & \multicolumn{2}{c|}{Baseline} & \multicolumn{2}{c|}{LumiUP} \\ \cline{2-5}
                                                         & 250 GeV & + 500 GeV  & 250 GeV & + 500 GeV \\
   \hline
   $g_{HWW}$                                    & 4.8\%       & 1.2\%      & 2.3\%     & 0.58\%     \\   
   $\Gamma_H$                                 & 11\%        & 5\%          & 5.4\%      & 2.5\%     \\            
   \hline
  \end{tabular}
  \caption{Expected precisions of total Higgs width and $HWW$ coupling for both baseline and luminosity upgrade (LumiUP) scenarios of ILC, at 250 GeV and 500 GeV, where the data at earlier stage is combined to later stage.}
\label{tab:HWWGamma}
  \end{center}
\end{table}

\acknowledgments
We would like to thanks all the members of the ILC physics subgroup and ILD optimization group for useful discussions, particularly to the software group T. Barklow, M. Berggren, A. Miyamoto and F. Gaede for preparing all the samples. This work is supported in part by the Creative Scientic Research Grant No. 18GS0202 of the
Japan Society for Promotions of Science (JSPS), the JSPS Grant-in-Aid for Science Research No. 22244031, and the JSPS Specially Promoted Research No. 23000002.

\bibliographystyle{hunsrt}
\bibliography{ref}

\end{document}